\documentclass[12pt]{article}

 \usepackage[vmargin=1.18in,hmargin=1.1in]{geometry}
 \usepackage{amsmath,centernot}
\usepackage{amsfonts}
\usepackage{amsthm}
\usepackage{natbib}
\usepackage{enumerate}
\usepackage{tikz}
\usepackage{setspace}
\usepackage{caption}
\usepackage{subcaption}

\usepackage[utf8]{inputenc}
\newcommand{\indep}{\perp \!\!\! \perp}

    \title{Alternative approaches for analysing\\ repeated measures data that are \\missing not at random} 
\author{Oliver Dukes$^1$, David Richardson$^2$,
and Eric Tchetgen Tchetgen$^1$\\
$^1$: University of Pennsylvania, Philadelphia, PA,  USA\\
$^2$: University of California Irvine,  Irvine,  CA,  USA}
\date{\today}

\begin{document}

\maketitle






\begin{abstract}
We consider studies where multiple measures of an outcome variable are collected over time, but some subjects drop out before the end of follow up. Analyses of such data often proceed under either a `last observation carried forward' or `missing at random' assumption. We consider two alternative strategies for identification; the first is closely related to the difference-in-differences methodology in the causal inference literature. The second enables correction for violations of the parallel trend assumption, so long as one has access to a valid `bespoke instrumental variable'. These are compared with existing approaches, first conceptually and then in an analysis of data from the Framingham Heart Study.  
\end{abstract}

\section*{Introduction}

In this paper, we consider the analysis of data where a variable of interest is measured at multiple time points, and at certain time points is subject to missing entries. We will consider in particular settings where there is drop-out without re-entry, such that once an individual has missing data at one time point, then data at all subsequent time points are also missing. Analyses with drop-out sometimes proceeds under a  Missing At Random (MAR) assumption, whereby one assumes that missingness depends on variables which are fully observed under a given missing data pattern. Such an assumption is convenient, as one can then employ methods such as multiple imputation or mixed models for repeated measures to remove potential biases; these are now widely available in standard software packages. 

Nevertheless, in many settings MAR may not be plausible. In that case, estimation and inference under an alternative Missing Not At Random (MNAR) assumption may be possible, albeit more challenging. Some examples of possible alternative assumptions are given in \citet{wu1988estimation,diggle1994informative,rotnitzky1997analysis, little2019statistical}. Much of the literature on analyses under MNAR mechanisms has focused on sensitivity analyses \citep{robins2000sensitivity,carpenter2007sensitivity}, rather than (or in addition to) reporting a single estimate and confidence interval for a parameter of interest. Typically, one fluctuates a sensitivity parameter in order to assess the consequences of violating a `reference' missing data assumption.

In this article, we will consider two alternative missing data assumptions, which may hold under certain MNAR mechanisms. Both harness the fact that the data include repeated measures from the same individuals. The first approach is an extension of the difference-in-differences (DiD) approach that is well known in the causal inference literature \citep{card1993minimum,lechner2011estimation}. This design makes a comparison between treatment arms in terms of a difference in the end of study outcome from baseline, and can recover a causal effect under a `parallel trends' assumption. We therefore extend this strategy to the missing data setting, and compare the parallel trends assumption both to MAR and to the controversial Last Observation Carried Forward (LOCF) assumption \citep{kenward2009last,lachin2016fallacies}. After this, we consider a distinct approach based on harnessing a `bespoke' Instrumental Variable (IV). This strategy was recently described for inferring causal effects in \citet{richardson2021bespoke} and \citet{dukes2022bespoke}. We show how a bespoke IV can be used to correct for violation of the parallel trends assumption, so long as the association between the bespoke instrument and the outcome is stable across time. 

Both the DiD and bespoke IV strategies are relatively straightforward to compute using existing software. Although we first describe them in a simple two-time point setting, they can be extended to longitudinal settings with many repeated measures. In Section \ref{sec_long} we describe their use for estimating the effect of a treatment in a clinical trial where outcomes are subject to monotone missingness. The DiD strategy we propose is closely related to the changes-in-changes approach described in \citet{ghanem2022correcting}; the proposal made here is generally simpler to implement, and we show how it can be extended to more general longitudinal data structures (those authors focus on a two-time point setting). Interestingly, methods closely related to the proposals described here have been implemented before in the analysis of clinical trials \citep{aisen2003effects,petersen2005vitamin,molnar2008does}, although the connection with the DiD design does not seem to have been made, nor have the assumptions been rigorously stated.  As far as we are aware, this is the first work to extend the bespoke IV strategy to missing data problems.

\section{Setting and existing methods}\label{sec_set}

To illustrate the methods and develop intuition for the assumptions, we will first focus on a simple setting, where data are collected on an outcome $Y$ at two consecutive time points. Consider the following data structure $\{Y_0,R,RY_1\}$ where $R$ is a missingness indicator, and $Y_t$, $t=0,1$ is an outcome of interest measured at time $t$. Suppose that there is interest in the full-data population mean of the final outcome $E(Y_1)$; then one can show that 
\begin{align*}
E(Y_1)
&=E(Y_1|R=1)-\{E(Y_1|R=1)-E(Y_1|R=0)\}Pr(R=0)
\end{align*}
where the conditional expectation $E(Y_1|R=0)$ is unknown. If we use the mean of $Y_1$ in the complete cases in place of $E(Y_1)$, then the bias is generally equal to
\[\{E(Y_1|R=1)-E(Y_1|R=0)\}Pr(R=0)\]
If data are ``missing completely at random'' (MCAR), then $R\indep Y_1$; in other words, the missingness indicator does not depend on the variable $Y_1$ which is subject to missingness. Under this assumption, it follows that the bias term given above is equal to zero. In contrast, if the data are MAR or MNAR, analysing data by restricting to the complete cases will generally lead to invalid results. 

When one has access to a baseline measurement $Y_0$, an approach that is sometimes leveraged to identify $E(Y_1)$ is LOCF. The underlying assumption can be phrased as 
\begin{align}\label{locf_as}
E(Y_1|R=0)=E(Y_0|R=0)
\end{align}
e.g. that the average of $Y_1$ in those with missing data is equal to the average of the earlier measurement in the same group. Under LOCF, one can show that $E(Y_1)$ can be identified as
\begin{align*}
E(Y_1)&=E(Y_1|R=1)-\{E(Y_1|R=1)-E(Y_0|R=0)\}Pr(R=0)
\end{align*}
One can show that the right hand side of the above is equal to 
\[E(Y_1|R=1)Pr(R=1)+E(Y_0|R=0)Pr(R=0)\]
which means that $E(Y_1)$ can straightforwardly be estimated using the observed data by substituting the missing outcomes with values of $Y_0$. Nevertheless, condition (\ref{locf_as}) is often deemed as implausible \citep{shao2003last,kenward2009last}, since it requires there to be no time trend in $Y_t$ in those with missing data. Despite criticism, this approach has been widely used in practice \citep{lachin2016fallacies}, likely due to its computational simplicity.

A second assumption often made that leverages the baseline $Y_0$ is MAR. In this context, the MAR assumption can be expressed as 
\[R\indep Y_1|Y_0\]
e.g. the missing data mechanism is independent of $Y_1$, within levels of the baseline measurement. This condition is often considered as a relaxation of the previous MCAR assumption. Repeating the previous decomposition of the mean of $Y_1$ whilst now conditioning on $Y_0$ gives
\begin{align*}
E(Y_1|Y_0)=&E(Y_1|R=1,Y_0)-\{E(Y_1|R=1,Y_0)-E(Y_1|R=0,Y_0)\}Pr(R=0|Y_0)
\end{align*}
and the MAR assumption implies that 
\[E(Y_1|R=1,Y_0)-E(Y_1|R=0,Y_0)\]
is equal to zero. Hence, one can identify $E(Y_1)$ by averaging the predictions $E(Y_1|R=1,Y_0)$ made in the  complete cases over the distribution of $Y_0$ in the full sample. More generally, under the MAR assumption, one can then estimate parameters of interest using maximum likelihood \citep{allison2012handling}, imputation approaches \citep{little2019statistical}, inverse probability weighting \citep{horvitz1952generalization,robins1994estimation}, or combinations of the above e.g. doubly robust methods \citep{robins1994estimation,robins1995analysis}. These all rely on the same missing data conditions for validity, but may require distinct parametric modelling assumptions (and possess differing statistical properties).

\section{Difference in differences}\label{sec_did}

The DiD assumption for missing data can be stated as : 
\begin{align}\label{pt}
E(Y_1-Y_0|R=1)=E(Y_1-Y_0|R=0)
\end{align}
The above condition assumes the trend in outcomes in the unobserved group equals the trend in the observed group. Such an assumption is illustrated in Figure \ref{did_locf_comp}(a), where the black line indicates the trend in average outcome values between time points in those with observed data. The dashed line is the trend in values in those with missing data; note that the point where the line crosses the $Y$ axis represents the measurement $E(Y_0|R=0)$ which is observed. However, because $E(Y_1|R=0)$ is unobserved, one would not observe this trend in those with $R=0$. 
Note that MCAR is not satisfied here, as $E(Y=1|R=1)\neq E(Y=1|R=0)$. Assumption (\ref{pt}) would nevertheless be satisfied, since the observed and unobserved trend curves are parallel, so that one can use the average difference in $Y_1-Y_0$ in the complete cases to predict $E(Y_1|R=0)$. Indeed, one can show that the full population mean $E(Y_1)$ is identified as 
\begin{align*}
E(Y_1)
&=E(Y_1|R=1)-\{E(Y_0|R=1)-E(Y_0|R=0)\}Pr(R=0).
\end{align*}
One can construct an estimator of $E(Y_1)$ based on the above expression, replacing the population expectations with the sample analogues. Alternatively, one can impute the missing outcomes with the baseline $Y_0$ plus the average change $Y_1-Y_0$ as estimated in the complete cases, before calculating the mean of the observed and imputed $Y_1$ values. Estimation of a standard error can be done using the nonparametric bootstrap (the same holds for other approaches discussed in this article).

The DiD assumption (\ref{pt}) is equivalent to the condition that 
\[E(Y_1|R=1)-E(Y_1|R=0)=E(Y_0|R=1)-E(Y_0|R=0)\]
\citep{sofer2016negative}. Note that this is no longer phrased in terms of the outcome trends, but rather concerns the bias in the complete case analysis. The assumption therefore states that this bias is exactly equal to the association between the baseline outcome and the missingness indicator, and therefore can be learnt from the data. In other words, the bias from restricting to complete cases is reflected by the baseline differences (in terms of the outcome) between those with and without missing data. 

Although as far as we are aware, the DiD strategy has received limited attention from missing data methodologists, versions of it nevertheless appear to have been applied in previous analyses of clinical trials. For example, in a trial comparing treatment with naproxen or rofecoxib versus placebo in patients with mild-to-moderate Alzheimer disease, the primary outcome was the Alzheimer Disease Assessment Scale-Cognitive (ADAS-Cog) subscale score (a measure of cognitive decline) at 12 months after randomisation \citep{aisen2003effects}. To impute missing outcome data, the estimated score change for the unobserved period in the relevant treatment arm complete cases was added to a participant's last observed ADAS-Cog score. The approach is described as `slope imputation', and is used in the primary analysis as an alternative to LOCF.

\subsection{Comparison with LOCF}

\begin{figure}
    \centering
     \begin{subfigure}[b]{0.5\textwidth}
    \includegraphics[width=0.95\textwidth]{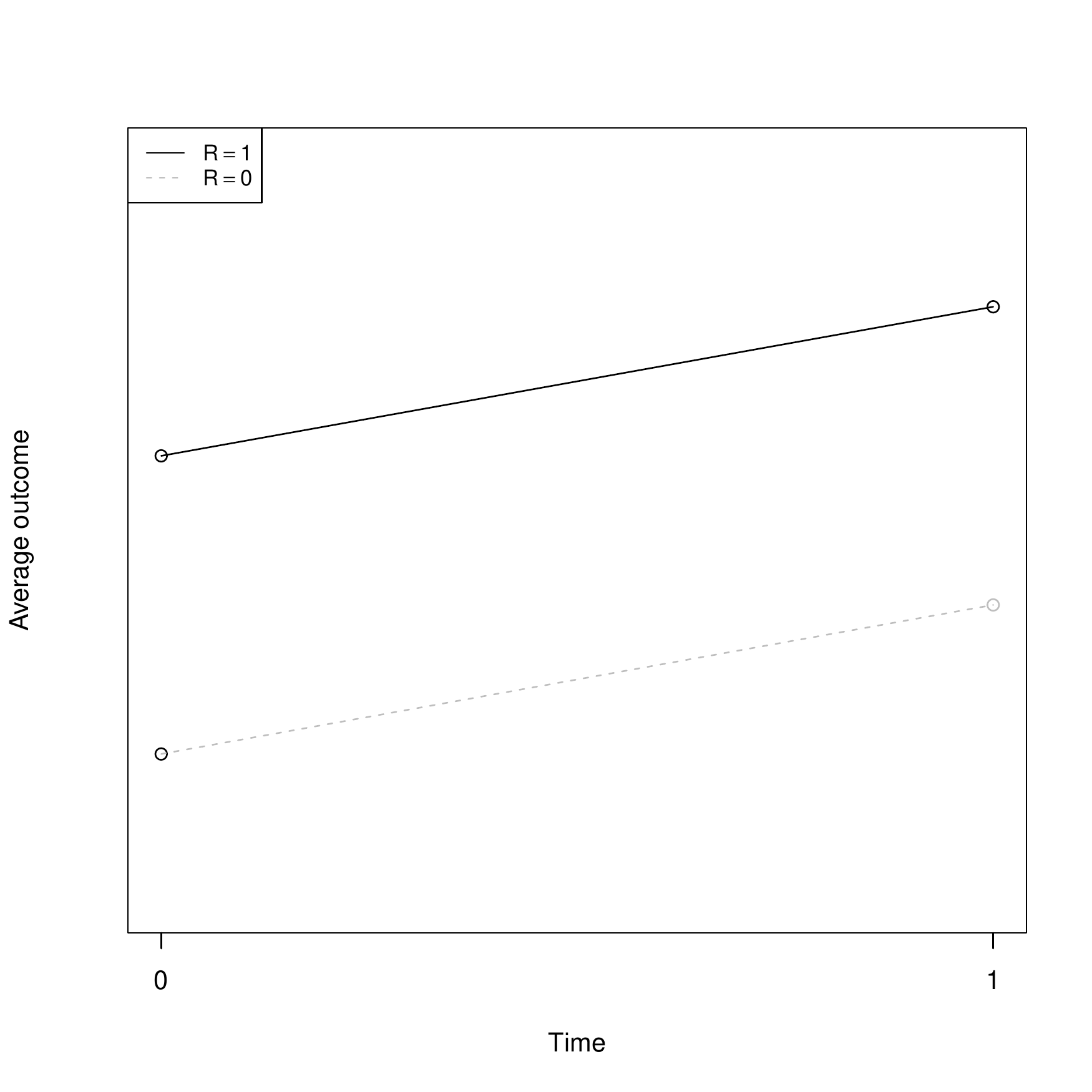}
             \caption{}
     \end{subfigure}\hfill
     \begin{subfigure}[b]{0.5\textwidth}
    \includegraphics[width=0.95\textwidth]{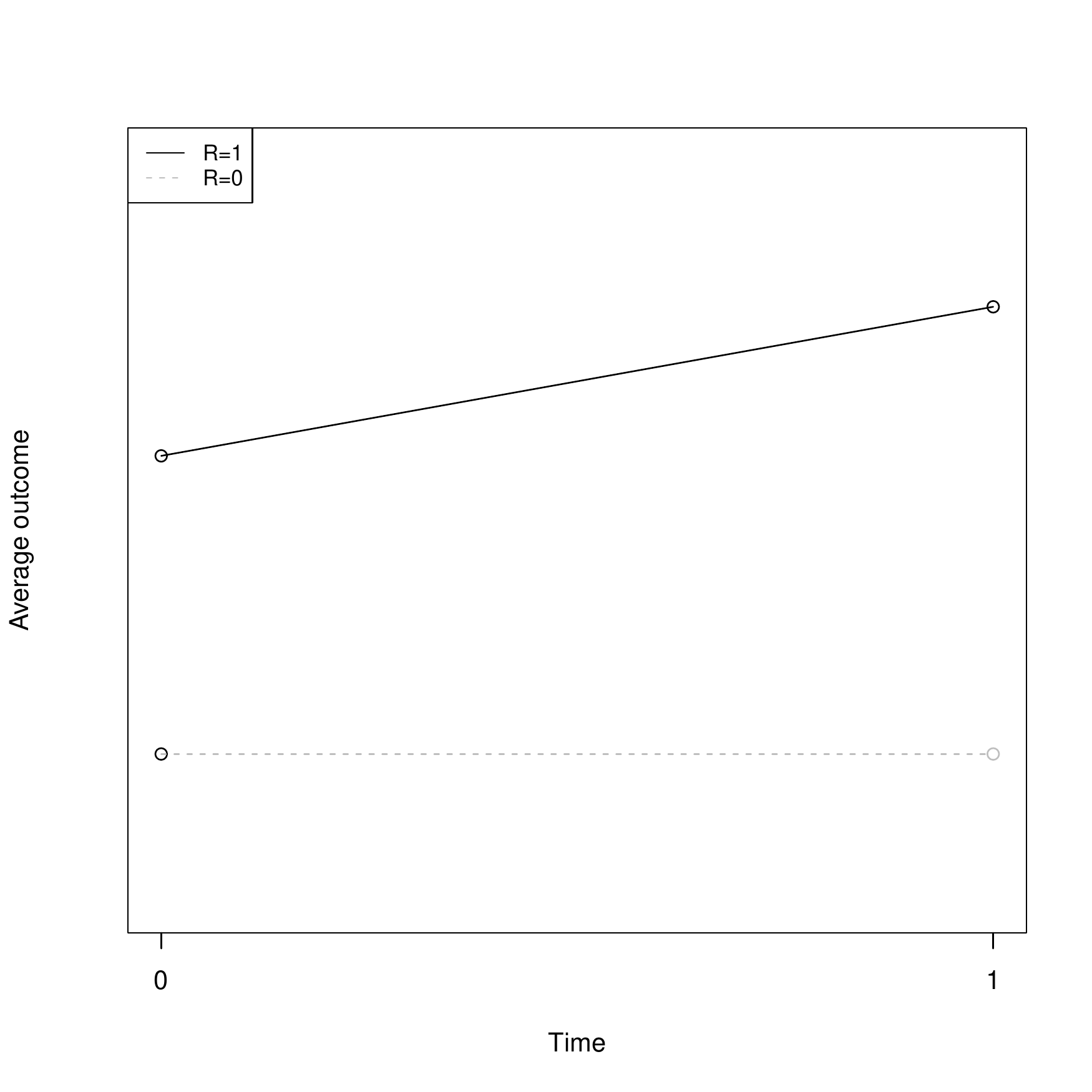}
                 \caption{}
\end{subfigure}
\caption{Two figures where the observed data is the same, but the missing data mechanism changes. The bold circles indicate the observed values of $E(Y_t|R=r)$ at $t=0$ and $t=1$ (if $R=1$). The faint circle indicates the unobserved $E(Y_1|R=0)$. Figure 1(a) represents a data generating mechanism where the parallel trends assumption (\ref{pt}) holds, whereas Figure 1(b) represents a setting where (\ref{pt}) is violated but (\ref{locf_as}) holds.}
\label{did_locf_comp}
\end{figure}

Assumption (\ref{pt}) is closely related to the LOCF assumption (\ref{locf_as}): it can be further rephrased as 
\[E(Y_1|R=0)=E(Y_0|R=0)+E(Y_1-Y_0|R=1)\]
in other words, that the mean of $Y_1$ in the group with missing data is equal to the baseline mean in that group (as in LOCF), plus the average trend in outcomes in the complete cases. The DiD assumption may often be more plausible than LOCF, if one believes that a time trend in the outcome is likely. However, the fact that $E(Y_1-Y_0|R=1)\neq 0$ (which can be checked from the data) does not necessarily support the DiD assumption over LOCF. This is because (\ref{pt}) is untestable; it may be that in truth there was no time trend in those with $R=0$. This is illustrated in Figure \ref{did_locf_comp}(b), where one sees that the unobserved time trend in those with $R=0$ is equal to zero. However, since there was a trend in the complete cases, a DiD analysis would erroneously use this trend to impute $Y_1$. In summary, a DiD analysis corrects the bias from LOCF only when the (untestable) parallel trends assumption holds.

One major criticism of LOCF is that even when the data is MCAR, it can still yield biased results. To make this concrete, suppose that that MCAR assumption $R\indep Y_1$ holds. The LOCF analysis then has a bias equal to \[\{E(Y_1|R=1)-E(Y_0|R=0)\}Pr(R=0)\] which is not equal to zero unless $E(Y_1|R=1)=E(Y_0|R=0)$. This is not implied either by MCAR or the LOCF assumption alone. On the other hand, note that the DiD analysis has a bias of \[\{E(Y_0|R=1)-E(Y_0|R=0)\}Pr(R=0);\] if one strengthens the previous MCAR condition to $R\indep (Y_1, Y_0)$, then it follows that this analysis in unbiased. In practice, it is difficult to think of settings where MCAR holds but the stronger condition does not, as $Y_1$ and $Y_0$ will usually be strongly correlated, so $R$ will typically depend on both (or neither).

\begin{figure}
    \centering
\begin{minipage}{0.5\textwidth}
	\begin{tikzpicture}[scale=2]
			\node[] (r) at (0, 2)   {$(a)$};
			\node[] (y0) at (0, 0)   {$Y_0$};
	\node[draw,circle]  (u) at (1, 1)   {$U$};
		\node[] (y1) at (2, 0)   {$Y_1$};
		\node[] (r) at (0, -1)   {$R$};
	\path[->,dashed] (u) edge node {} (y0);
		\path[->,dashed] (u) edge node {} (y1);
		\path[->] (y0) edge node {} (r);
				\path[->] (y0) edge node {} (y1);

	\end{tikzpicture}
\end{minipage}%
\begin{minipage}{.5\textwidth}
	\begin{tikzpicture}[scale=2]
				\node[] (r) at (0, 2)   {$(b)$};
			\node[] (y0) at (0, 0)   {$Y_0$};
	\node[draw,circle]  (u) at (1, 1)   {$U$};
		\node[] (y1) at (2, 0)   {$Y_1$};
		\node[] (r) at (0, -1)   {$R$};
	\path[->,dashed] (u) edge node {} (y0);
		\path[->,dashed] (u) edge node {} (y1);
		\path[->,dashed] (u) edge node {} (r);

	\end{tikzpicture}
\end{minipage}%
\\
\begin{minipage}{.5\textwidth}
	\begin{tikzpicture}[scale=2]
			\node[] (r) at (0, 2)   {$(c)$};
			\node[] (y0) at (0, 0)   {$Y_0$};
	\node[draw,circle]  (u) at (1, 1)   {$U$};
		\node[] (y1) at (2, 0)   {$Y_1$};
		\node[] (r) at (0, -1)   {$R$};
	\path[->,dashed] (u) edge node {} (y0);
		\path[->,dashed] (u) edge node {} (y1);
		\path[->] (y0) edge node {} (r);
						\path[->,dashed] (u) edge node {} (r);
		\path[<->,dashed] (y0) edge [bend left]  node {} (y1);


	\end{tikzpicture}
\end{minipage}%
\begin{minipage}{.5\textwidth}
	\begin{tikzpicture}[scale=2]
				\node[] (r) at (0, 2)   {$(d)$};
			\node[] (y0) at (0, 0)   {$Y_0$};
	\node[draw,circle]  (u) at (1, 1)   {$U$};
		\node[] (y1) at (2, 0)   {$Y_1$};
		\node[] (r) at (0, -1)   {$R$};
	\path[->,dashed] (u) edge node {} (y0);
		\path[->,dashed] (u) edge node {} (y1);
		\path[->,dashed] (u) edge node {} (r);
		\path[->] (y0) edge node {} (y1);
		\path[<->,dashed] (y0) edge [bend left]  node {} (y1);

	\end{tikzpicture}
\end{minipage}

\caption{Four causal diagrams illustrating different missing data mechanisms.}
\label{standard_proxy_med}
\end{figure}

\subsection{Comparison with MAR}

This assumption is also distinct from MAR, as it neither implies it nor is implied by it. The comparison is similar to the distinction between the parallel trends assumption and conditional mean exchangeability (given the baseline outcome) in the causal inference literature. To see a setting where MAR may be preferable, consider the causal diagram in Figure \ref{standard_proxy_med}(a); drop-out is directly influenced by the baseline outcome $Y_0$, which is associated with the outcome of interest via an unmeasured latent variable $U$. It follows from the diagram that conditioning on $Y_0$ renders $Y_1$ and the missingnness indicator independent. However, in Figure \ref{standard_proxy_med}(b), drop-out is influenced instead by the unmeasured $U$, rather than $Y_0$; in this case there would still be residual dependence between $R$ and $Y_1$. This figure corresponds to an MNAR mechanism. 

Still focusing on Figure  \ref{standard_proxy_med}(b), suppose that the following linear models hold:
\begin{align*}
E(Y_t|R,U)&=\alpha_t+\beta U\\
E(U|R)&=\delta+\gamma R
\end{align*}
The first equation encodes the restriction that $R$ is independent of $Y_1$ given $U$; importantly, the association between $U$ and the outcome is also constant across time points. Then by averaging out $U$, we have 
\begin{align*}
E(Y_t|R)&=\alpha_t+\beta\{\delta+\gamma R\}
\end{align*}
so although $E(Y_1|R)$ and $E(Y_0|R)$ both depend on $R$, we see that $E(Y_1-Y_0|R)=\alpha_1-\alpha_0$ does not. Hence in this setting, whilst an analysis under MAR may be misleading, the DiD assumption (\ref{pt}) holds. We note that (\ref{pt}) can be used as a primitive assumption, and one does not need the previous system of linear equations to hold; nor does one need to make assumptions about the stationarity of the association between $Y_t$ and $U$. Nevertheless, this example may help to reason about settings where this assumption is plausible and others may fail. 

Figures  \ref{standard_proxy_med}(c) and  \ref{standard_proxy_med}(d) illustrate other settings where MAR would generally fail, but a DiD analysis may be unbiased so long as (\ref{pt}) holds. A property shared by Figures  \ref{standard_proxy_med}(b)- \ref{standard_proxy_med}(d) is that $Y_0$ is not a common cause of $R$ and $Y_1$, but rather causes either one or the other (or neither). In these cases $Y_0$ is not itself responsible for the missing data bias. This is a distinction with the MAR analysis, where $Y_0$ is conditioned on to close the `backdoor path' between $R$ and $Y_1$ \citep{pearl1995causal}. 

To summarise, the MAR and DiD assumptions are in general distinct and non-overlapping. There are however settings where both may hold, although these are perhaps artificial. One setting where both would coincide is if we were to assume that the following linear model holds:
\begin{align}\label{mar_did}
E(Y_1|R,Y_0)=\alpha^*_1+Y_0 
\end{align}
Note that this implies the MAR assumption as the right hand side does not depend on $R$; however, stronger parametric restrictions are made, namely that the coefficient in the linear model for $Y_0$ is exactly equal to 1. One can show that this model also implies (\ref{pt}). We emphasise that in a DiD analysis, one does not require (\ref{mar_did}) to hold; rather, the parallel trends assumption (\ref{pt}) and MAR coincide under (\ref{mar_did}). 

\section{Bespoke instrumental variables}\label{sec_bsiv}

In certain cases, the parallel trends assumption (\ref{pt}) will not be plausible. For example, in the trial evaluating different treatments for Alzheimer Disease, it may be that the change in ADAS-Cog scores in the active participants who dropped out tended to be less pronounced than the change observed in the complete cases. Assumption (\ref{pt}) may fail when there are unmeasured factors which are common causes of $Y_0$ and $R$ which differ from the unmeasured common causes of $Y_1$ and $R$. The bias of the DiD estimate of $E(Y^1)$ can be shown to equal 
\[\{E(Y_1-Y_0|R=1)-E(Y_1-Y_0|R=0)\}Pr(R=0).\]
It can be easily verified that this expression equals zero when either (\ref{pt}) holds or $Pr(R=0)=0$ (there is no missing data).  In what follows, we will look at how collecting data on an additional auxiliary variable $Z$ can help to relax (\ref{pt}). This comes at the cost of some alternative assumptions made between the outcome process and $Z$, which in certain cases may be more plausible.

Suppose a binary variable $Z$ satisfies the following key conditions
\begin{itemize}
    \item (Stability of association)
    \begin{align}\label{er}
    E(Y_1|Z=1)-E(Y_1|Z=0)=E(Y_0|Z=1)-E(Y_0|Z=0)
    \end{align}
    \item (Bias homogeneity)
    \begin{align}\label{nsm}
    &E(Y_1-Y_0|Z=1,R=1)-E(Y_1-Y_0|Z=1,R=0)\nonumber\\&=E(Y_1-Y_0|Z=0,R=1)-E(Y_1-Y_0|Z=0,R=0)
    \end{align}
    \item (Predictive of missingness)
    \begin{align}\label{pos}
        Pr(R=1|Z=1)-Pr(R=1|Z=0)\neq 0
    \end{align}
\end{itemize}
The first condition requires that the association between $Y_1$ and $Z$ in the full population is the same as the association between $Y_0$ and $Z$. Similar to (\ref{pt}), this assumption cannot be tested, since we do not have access to $Y_1$ for everyone. It requires choosing a covariate for which the association with the outcome is likely to be stable over time, and is not subject to temporal trends. 

Assumption (\ref{er}) is a type of `exclusion restriction', since we exclude an association between the mean of $Y_1-Y_0$ and $Z$. It is related to the assumption in the causal inference literature that an instrumental variable (IV) cannot have a direct causal effect on the outcome. \citet{tchetgen2017general} previously considered the setting with an outcome missing not at random, and showed how an IV for missing outcome can be used to correct for bias. They relied on the exclusion restriction that $E(Y_1|Z=1)=(Y_1|Z=0)$ e.g. that it is not associated with the outcome in the full data. However, in practice it might be hard to find variables for which this assumption is satisfied. Our assumption is close to Tchetgen Tchetgen and Wirth's, but is made after `differencing' the outcome; hence we can allow for $Z$ to be associated with $Y_1$ in the full data, so long as its association matches that with $Y_0$. The link with the causal inference literature has lead to this type of approach being described a `bespoke IV' strategy, where a standard confounder $Z$ measured at baseline is rendered as an IV \citep{richardson2021bespoke,dukes2022bespoke}. We use the same terminology here. 

Assumption (\ref{nsm}) states that the bias in the DiD analysis is the same for those with $Z=1$ as for those with $Z=0$. Since we do not know the selection bias, this is also not an assumption that can be empirically verified. Hence, the assumption (\ref{nsm}) allows for deviations from the parallel trends assumption (\ref{pt}), so long as these deviations do not depend on $Z$. Assumption (\ref{pos}) requires $Z$ to be predictive of missingness; unlike the other assumptions, this condition can be verified empirically. Therefore, a candidate $Z$ could potentially be a covariate that would usually be conditioned on in a MAR analysis, that is associated with both the outcome and the probability of missingness. A good choice of $Z$ is additionally required to have a stable association with the outcome over time, and for the DiD bias to be the same within each of its levels. It may be that assumption (\ref{nsm}) is the hardest to justify on scientific grounds; assumption (\ref{er}) could potentially be motivated by external data or existing scientific knowledge.

In the Appendix, we show that assumptions (\ref{er})-(\ref{pos}) are sufficient to identify $E(Y_1)$. This result depends on the fact that we are able (under these assumptions) to represent the observed data as 
\begin{align*}
E(Y_1-Y_0|Z,R=1)&=E(Y_1-Y_0)+\{E(Y_1-Y_0|R=1)-E(Y_1-Y_0|R=0)\}Pr(R=0|Z)
\end{align*}
Notice that the left hand side is a conditional expectation that involves only the observed data. The right hand side involve two quantities that are unknown: $E(Y_1-Y_0)$ and $E(Y_1-Y_0|R=1)-E(Y_1-Y_0|R=0)$ (one can straightforwardly calculate $Pr(R=0|Z)$ without using data on $Y_1$ or $Y_0$). This is tantamount to a regression problem, where $Y_1-Y_0$ is the outcome, $Pr(R=0|Z)$ is the regressor, and we restrict only to the complete cases; $E(Y_1-Y_0)$ and $E(Y_1-Y_0|R=1)-E(Y_1-Y_0|R=0)$ are the intercept and slope parameters respectively. This motivates the following procedure for estimating $E(Y_1)$:
\begin{enumerate}
  \item Obtain the predictions $Pr(R=1|Z)$ using e.g. logistic regression.
    \item Fit a linear regression model using the complete cases, with $Y_1-Y_0$ as the outcome, $Pr(R=1|Z)$ as a predictor, and $-E(Y_0)$ as an offset:
    \begin{align}\label{reg_mod}
    E(Y_1-Y_0|Z,R=1)=-E(Y_0)+\tau_0+\tau_1 Pr(R=0|Z)
    \end{align}
    \item Estimate $E(Y_1)$ as the the estimated value of $\tau_0$. 
\end{enumerate}
In the regression model (\ref{reg_mod}), the parameters $\tau_0$ and $\tau_1$ correspond with $E(Y_1)$ and the DiD bias respectively. 





\section{Data analysis: Framingham Heart Study}

In order to illustrate the aforementioned methods we considered a subsample of data from the Framingham Heart Study. This is a landmark prospective study in cardiovascular epidemiology, conducted using individuals from Framingham, Massachusetts; participants are followed up every 2 to 8 years. Our sample included measurements on clinical examination data from when the first group of subjects were enrolled in 1956, as well as measurements on the same variables from their third examination in 1968. Out of the 5,209 participants who initially enrolled, the subsample consisted of 400 individuals, who were sampled such that the dataset was balanced in terms of sex and smoking status; the data set was obtained from the `sur' package in R \citep{weinberg2020statistics}. 

Our parameter of interest was average systolic blood pressure (SBP) at the third examination; this was missing for 92 individuals (23\%). Complete data was collected on SBP at the first examination. We compare several approaches; an analysis under a MCAR assumption, based on comparing average outcomes in the complete cases; a LOCF analysis, where missing outcomes in each arm are imputed using baseline outcome; a MAR analysis, where baseline outcomes are assumed to render missingness and outcomes independent; a DiD analysis, based on (\ref{pt}); and a bespoke IV analysis. For the final approach, we chose age as a bespoke instrument. Age was highly correlated with missingness; participants with missing data were on average 4.49 years older than the complete cases (95\% CI:  2.56, 6.41). We implemented two bespoke IV analyses, where we varied the specification of the model for $Pr(R=1|Z)$. The first approach involved a logistic regression with a linear term for age (BSIV), whereas the second included an additional quadratic term in the model (BSIV-Q) 

\begin{figure}
    \centering
    \includegraphics[width=0.5\textwidth]{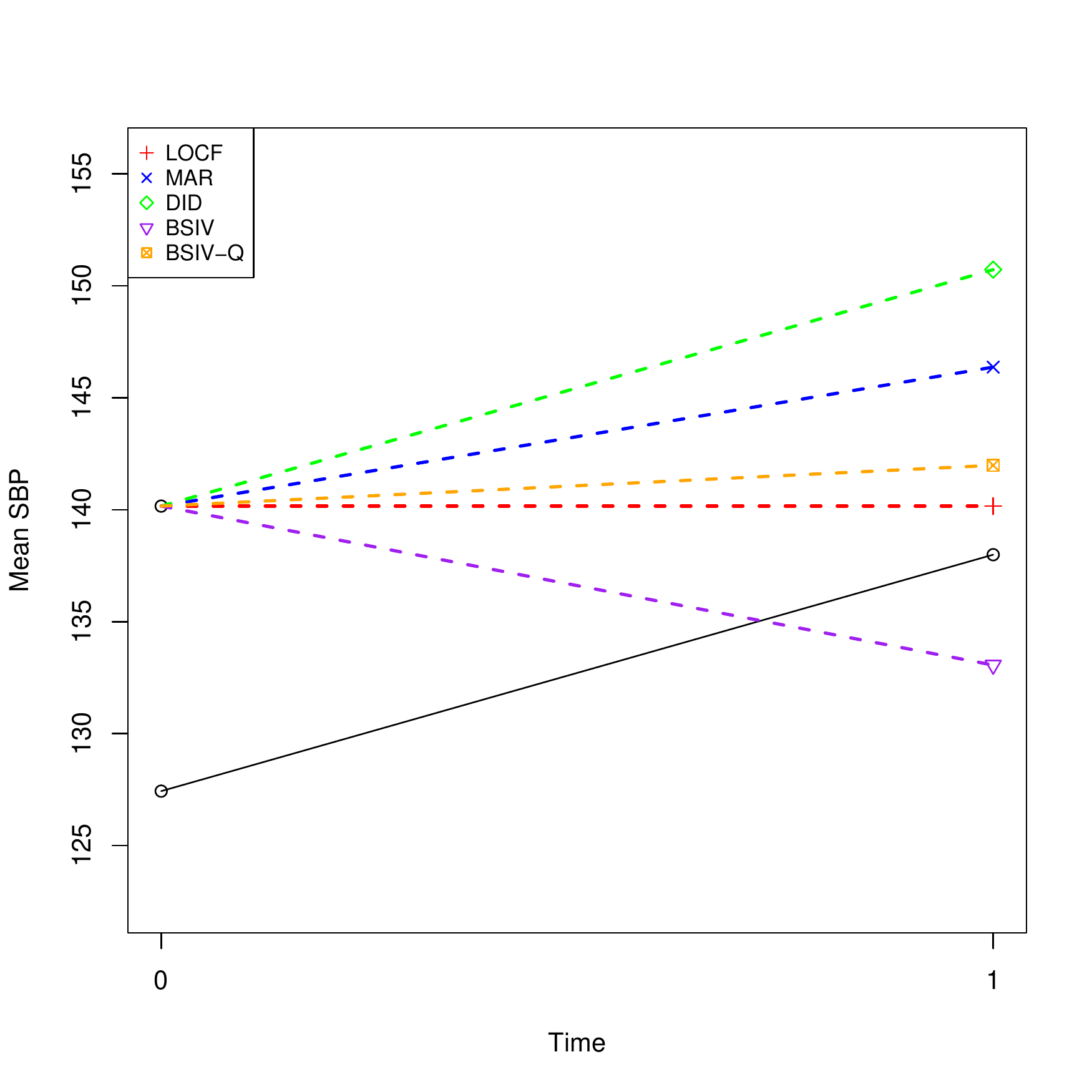}
\caption{Estimates of mean SBP from the Framingham Heart Study for those with and without missing data at third examination.  The black circles and line describe the trajectory for the complete cases ($R=1$). The other lines indicate the different trajectories for those with $R=0$, as the missing data assumptions are varied.}
\label{fhs_plot}
\end{figure}

In Figure \ref{fhs_plot}, we give a graphical illustration of how the different approaches estimate $E(Y_1|R=0)$. The bold black line indicates the change in outcomes between first and third examination in the complete cases. It is evident that baseline values tended to be higher in those who then had missing data at follow up. One can see that the LOCF approach would give similar estimates of $E(Y_1|R=1)$ as an MCAR analysis, since mean SBP at third examination is close to the mean of SBP at baseline in those with missing data at follow up. The MAR and DiD analysis suggest that SBP would tend to increase over time in those with $R=0$; the DiD approach yields the highest estimate of $E(Y_1|R=0)$, and may be considered most plausible if one expects that SBP typically increases in individuals over time. A test of the null hypothesis that $E(Y_1-Y_0|R=1)$ equals zero yielded a $p$-value $<0.01$, suggesting that (under a parallel trends assumption) a LOCF analysis is biased. The bespoke IV analysis yielded the lowest estimate of $E(Y_1|R=0)$, illustrating how this approach may sometimes yield results that contradict those of the DiD methods. Here, bespoke IV assumes that the trend in SBP is the same (in the full population) does not depend on age; this would be violated if in truth older patients tended to see a more dramatic rise in SBP compared with younger patients. It may be in this case that the parallel trends assumption is more plausible. However, including a quadratic term led to results more in line with those for the other methods, suggesting some sensitivity to model specification.  

The estimates of mean SBP at third examination, along with nonparametric bootstrap 95\% confidence intervals, are provided in Table \ref{fhs_tab}. The point estimates are fairly close in magnitude, likely due to the relatively modest proportion of missing values. One can see that the BSIV intervals are wider than for the other approaches, overlapping with those from the MAR and DiD analyses; this should caveat the observed differences in the point estimates. 

\begin{table}[ht]
\centering
\caption{Estimates of mean SBP at third examination under different missing data assumptions in the Framingham Heart Study. 95\% CI: 95\% Confidence Interval (based on the nonparametric bootstrap - 2,000 bootstrap samples used)}
\begin{tabular}{lll}
  \hline
Analysis & Estimate & $95\%$ CI \\ 
  \hline
MCAR & 137.99 & (135.48,140.51) \\ 
  LOCF & 138.49 & (136.23,140.76) \\ 
  MAR & 139.92 & (137.31,142.53) \\ 
  DiD & 140.92 & (138.32,143.52) \\ 
  BSIV & 136.86 & (129.5,144.22) \\ 
 BSIV-Q & 138.91 & (131.63,146.19) \\ 
   \hline
\end{tabular}
\label{fhs_tab}
\end{table}

\section{Longitudinal clinical trials with monotone missing data}\label{sec_long}

\subsection{Difference-in-Differences}

Having introduced the DiD and bespoke IV strategies for handling a single missing outcome, we will now extend our reasoning for clinical trials subject to monotone missing outcome data. Let $A$ now denote a binary treatment variable of interest. Then we will consider a clinical trial subject to drop-out, where repeated measurements have been collected of the outcome at $T$ time points, in addition to a baseline measurement $Y_0$. Then the observed data takes the form $\{Y_0,A,R_1,R_1Y_1,...,R_T,R_TY_T\}$; in what follows, let $\bar{R}_t=\{R_1,...,R_t\}$. We will assume a monotone missing data structure, such that once an individual drops out of the study, no further data is collected on them; in other words, if $R_j=0$ at time $t=j$ for a participant, then $R_k=0$ for all $k>j$.
Interest is in the causal contrast \[E(Y^1_T-Y^0_T)\] where $Y_T^a$ is the potential outcome at time $T$ if an individual were given treatment $A=a$. By virtue of randomisation, this quantity is equal to $E(Y_T|A=1)-E(Y_T|A=0)$. However, due to the presence of missing data, one can still not readily compute this from the observed data without further assumptions.  In what follows, for simplicity we will consider identification of $E(Y^1)=E(Y_T|A=1)$; results along the same lines follow for $E(Y^0)$.  


One can show that the generalised version of the DiD parallel trend assumption can be given as 
\begin{align}\label{long_pt}
E(Y_T-Y_{t-1}|\bar{R}_{t}=1,A=1)=E(Y_T-Y_{t-1}|R_{t}=0,\bar{R}_{t-1}=1,A=1)
\end{align}
for $t=1,...,T$. Interpretation of this assumption is more subtle than in the previous sections, because at each time $t<T$, the group for whom $\bar{R}_{t}=1$ will include a mix of the complete cases $(\bar{R}_T=1)$ and individuals who subsequently drop-out. As a result, neither the left (unless $t=T$) nor the right hand side of (\ref{long_pt}) can be directly calculated from the observed data. 

\begin{figure}
    \centering
    \includegraphics[width=0.5\textwidth]{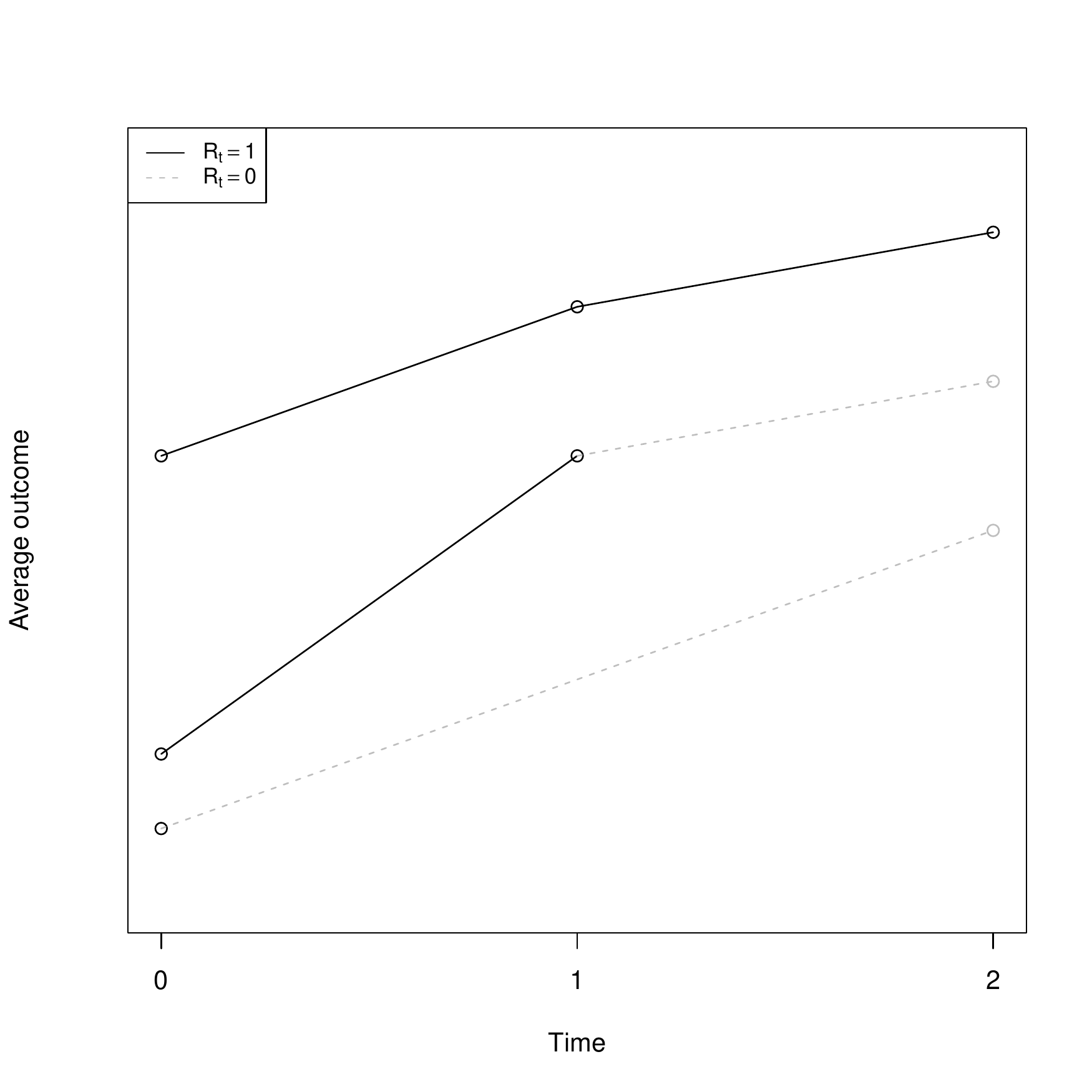}
\caption{The top line indicates the outcome trajectory in the complete cases. The middle  line indicates the trajectory in participants for whom $R_2=1, R_1=1$ under (\ref{pt}); the full black (dashed grey) line represents the observed (unobserved) data. The bottom line represents the  trajectory for participants for whom data was missing subsequent to baseline under (\ref{pt}). Here, $Pr(R_2=1|R_1=1,A=1)=Pr(R_2=0|R_1=1,A=1)=0.5$.}
\label{sdid}
\end{figure}

To develop some understanding of this assumption, consider first the group of patients with missing data only at the very end of follow up. For the sake of simplicity, we let $T=2$. For these individuals, the above restriction implies that 
\begin{align}\label{init}
E(Y_2|R_2=0,R_1=1,A=1)=&E(Y_1|R_2=0,R_1=1,A=1)+E(Y_2-Y_{1}|\bar{R}_2=1,A=1)
\end{align}
such that one can impute $Y_T$ in these individuals based on adding the trend $Y_2-Y_1$ (as observed in the complete cases) to their outcomes at time 1. So far, this straightforwardly follows what was described in Section \ref{sec_did}. For individuals who drop-out before $Y_1$ is measured ($R_1=0$), one can show that assumption (\ref{long_pt}) implies that 
\begin{align*}
&E(Y_2|R_1=0,A=1)\\&=E(Y_0|R_1=0,A=1)\\
&\quad+E(Y_2-Y_0|\bar{R}_{2}=1,A=1)Pr(R_{2}=1|R_1=1,A=1)\\
&\quad+\{E(Y_2-Y_1|\bar{R}_{2}=1,A=1)+E(Y_1-Y_0|R_2=0,R_1=1,A=1)\}Pr(R_2=0|R_1=1,A=1)
\end{align*}
Hence we impute their outcomes using the baseline $Y_0$, plus a weighted average of the observed  and \textit{predicted} trends $Y_2-Y_0$. The parallel trends assumption allows us to predict the trend in those for whom $Y_1$ was observed but $Y_2$ was not as $E(Y_2-Y_1|\bar{R}_{2}=1,A=1)+E(Y_1-Y_0|R_2=0,R_1=1,A=1)$, where information is again borrowed from the complete cases. Figure \ref{sdid} displays some potential observed and predicted trajectories in the complete cases (top), those with $R_2=0, R_1=1$ (middle) and $R_2=R_1=0$ (bottom). In particular, the bottom dashed line illustrates the predicted trajectory in the group with data missing subsequent to baseline; it is a weighted average of the trajectories in the two other missing data groups. 

There are multiple ways to implement the sequential DiD approach. In the Appendix we show that under assumption (\ref{long_pt}), 
\begin{align*}
&E(Y_T|A=1)\\&=E(Y_T|\bar{R}_T=1,A=1)\\
&\quad-\sum^{T}_{t=1}\{E(Y_{t-1}|\bar{R}_t=1,A=1)-E(Y_{t-1}|R_t=0,\bar{R}_{t-1}=1,A=1)\}Pr(R_t=0|\bar{R}_{t-1}=1,A=1)
\end{align*}
where the right hand side involves terms that can be readily computed from the data. Loosely, we subtract an estimate of the bias (using the immediate outcomes across different time points) from the na\"ive complete case analysis. Alternatively, one can impute the outcomes working backwards in time using predicted trajectories as hinted at in the previous paragraph. A third way to implement the method is via the representation
\[E(Y_T|A=1)=E(Y_0|A=1)+\sum^{T}_{t=1}E(Y_t-Y_{t-1}|\bar{R}_t=1,A=1)\]
such that one can calculate the average difference $Y_t-Y_{t-1}$ in those with $R_t=1$ across different time points and then sum them, adding on $E(Y_0|A=1)$.


An alternative assumption to (\ref{long_pt}) would be 
\begin{align}\label{long_pt_alt}
E(Y_T-Y_{t-1}|\bar{R}_{T}=1,A=1)=E(Y_T-Y_{t-1}|R_{t}=0,\bar{R}_{t-1}=1,A=1)
\end{align}
where in the conditional expectation on the left hand side, we now condition on being a complete case. Note that the left hand side can now be readily computed from the data, which was not the case with (\ref{long_pt}). An analysis based on (\ref{long_pt_alt}) would impute missing outcomes using the slopes based on the complete cases alone, rather than harnessing additional information from individuals who drop-out at later time points. In general, we view assumption (\ref{long_pt}) as more plausible; nevertheless, it is  (\ref{long_pt_alt}) that appears to have been applied before \citep{aisen2003effects}, perhaps because it is more intuitive and can be easily implemented. Referring back to Figure \ref{sdid}, under (\ref{long_pt_alt}), the bottom dashed line would be flatter, as it would be based entirely on the trajectory in the complete cases, rather than borrowing information from the steeper trajectory in participants with partially missing data.

A weakness of analyses developed either under assumptions (\ref{long_pt}) and (\ref{long_pt_alt}) is that they may lack plausibility when there are big gaps in time between measurement of $Y_t$ and earlier time points. We note that there is a growing literature on the failure of parallel trends in the causal inference literature on the DiD design \citep{roth2022s}, and many of these ideas may also be useful in this context.

\subsection{Bespoke instrumental variables}

The bespoke instrumental variable method can also be extended to the longitudinal set-up. We will now assume that one has access to a time-dependent variable $Z=\{Z_1,...,Z_T\}$, where we assume the temporal ordering $\{Z_t,R_t,R_tY_t\}$ for all $t=1,...,T$. We also need to extend the three previous bespoke IV conditions from Section \ref{sec_bsiv}:

\begin{itemize}
    \item (Stability of association) For all $t=0,...,T$,
    \begin{align}
    &E(Y_T|\bar{R}_{t-1}=1,Z_{t}=1,\bar{Z}_{t-1},A=1)-E(Y_T|\bar{R}_{t-1}=1,Z_{t}=0,\bar{Z}_{t-1},A=1)\nonumber\\
    &=E(Y_{t-1}|\bar{R}_{t-1}=1,Z_{t}=1,\bar{Z}_{t-1},A=1)-E(Y_{t-1}|\bar{R}_{t-1}=1,Z_{t}=0,\bar{Z}_{t-1},A=1)\label{er_l}
    \end{align}
    \item (Bias homogeneity) For all $t=0,...,T$,
    \begin{align}\label{nsm_l}
    &E(Y_T-Y_{t-1}|\bar{R}_{t}=1,Z_{t}=1,\bar{Z}_{t-1},A=1)\nonumber\\&\quad-
    E(Y_T-Y_{t-1}|R_{t}=0,\bar{R}_{t-1}=1,Z_{t}=1,\bar{Z}_{t-1},A=1)\nonumber\\
    &=E(Y_T-Y_{t-1}|\bar{R}_{t}=1,Z_{t}=0,\bar{Z}_{t-1},A=a)\nonumber\\&\quad-
    E(Y_T-Y_{t-1}|R_{t}=0,\bar{R}_{t-1}=1,Z_{t}=0,\bar{Z}_{t-1},A=a)
    \end{align}
    \item (Predictive of missingness) For all $t=0,...,T$,
    \begin{align}\label{pos_l}
        &Pr(R_t=1|\bar{R}_{t-1}=1,Z_{t}=1,\bar{Z}_{t-1},A=a)-Pr(R_t=1|\bar{R}_{t-1}=1,Z_{t}=0,\bar{Z}_{t-1},A=a)\nonumber\\&\neq 0
    \end{align}
\end{itemize}
Conditions (\ref{er_l}), (\ref{nsm_l}) and (\ref{pos_l}) generalise (\ref{er}), (\ref{nsm}) and (\ref{pos}) in the same way that the sequential parallel trend assumption (\ref{long_pt}) generalises (\ref{pt}). As in that case, for any $t$ up to $T-1$, the assumptions refer to a subgroup of participants, some of whom will subsequently drop-out. Assumption (\ref{nsm_l}) can be see as a weakening of the sequential parallel trends assumption (\ref{long_pt}), where parallel trends at time $t$ is allowed to be violated so long as the violation does not depend on $Z_t$. 

In the Appendix, we show that when assumptions (\ref{er_l})-(\ref{pos_l}) hold, 
\begin{align*}
&E(Y_T|A=1)\\
&\quad+\sum^{T}_{t=1}\{E(Y_{t}-Y_{t-1}|\bar{R}_t=1,\bar{Z}_{t-1}=\bar{z}_{t-1},A=1)-E(Y_{t}-Y_{t-1}|R_t=0,\bar{R}_{t-1}=1,\bar{Z}_{t-1}=\bar{z}_{t-1},A=1)\}\\&\quad \times Pr(R_t=1|\bar{R}_{t-1}=1,\bar{Z}_{t}=\bar{z}_{t},A=1)\\
&=E(Y_0|A=1)+\sum^{T}_{t=1}E(Y_{t}-Y_{t-1}|\bar{R}_t=1,\bar{Z}_t=\bar{z}_t,A=1)
\end{align*}
The left hand side of the equality involves the unknown $E(Y_T|A=1)$ and  $E(Y_{t}-Y_{t-1}|R_t=0,\bar{R}_{t-1}=1,\bar{Z}_{t-1}=\bar{z}_{t-1},A=1)$, whereas the right hand side involves quantities that can be obtained from the data.
Below, we use this result as a recipe for implementation:
\begin{enumerate}
    \item For each $t=1,...,T-1$, fit a model for 
    \[E(Y_{t}-Y_{t-1}|\bar{R}_t=1,\bar{Z}_t,A=1)\]
    by regressing $Y_{t}-Y_{t-1}$ on $\bar{Z}_t$ in those with $\bar{R}_t=1$ and $A=1$.
    \item In the complete cases ($\bar{R}_T=1$), add together the model predictions in the following way:
    \[\omega(\bar{Z}_{T-1})=E(Y_0|A=1)+\sum^{T-1}_{t}E(Y_{t}-Y_{t-1}|\bar{R}_t=1,\bar{Z}_t,A=1)\]
    \item For each $t=1,...,T$, fit a regression model for the probability of remaining in the risk set:
    \[Pr(R_t=1|\bar{R}_{t-1}=1,\bar{Z}_{t},A=1)\]
    \item Fit the following regression model in the complete cases in the treated arm:
    \begin{align*}
    &E(Y_T-Y_{T-1}|\bar{R}_{T}=1,\bar{Z}_t,A=1)\\
    &=-\omega(\bar{Z}_{T-1})+\tau_0+\sum^{T}_{t}\tau_{t}Pr(R_t=0|\bar{R}_{t-1}=1,\bar{Z}_{t}=\bar{z}_{t},A=1)
    \end{align*}
    where $-\omega(\bar{Z}_t)$ is included as an offset. Then $\tau_0$ is an estimate of $E(Y_T|A=1)$.
\end{enumerate}
At the final stage, one can also allow for interactions between the regressor $Pr(R_t=0|\bar{R}_{t-1}=1,\bar{Z}_{t}=\bar{z}_{t},A=1)$ and $\bar{Z}_{t-1}$ for any $t=1,...,T$, but the interaction cannot involve $Z_t$ by assumption (\ref{nsm_l}). 

\section{Discussion}

In this approaches, we have discussed two alternatives to the MAR assumption as a means of obtaining identification in settings with missing data.  The first approach extends the DiD design from causal inference, via a missing data parallel trends type assumption. The second approach allows for a violation of this assumption, and is valid under alternative conditions. It requires collecting data on an auxiliary covariate, but such covariate data is often collected in order to make the MAR assumption more plausible. We have limited our attention to settings with monotone missing data. Non-monotone missingness structures pose additional challenges for identification, which are beyond the scope of this article \citep{robins1997non}.

In DiD analyses, one often makes a parallel trends assumptions \textit{conditional on measured covariates} \citep{abadie2005semiparametric}. This enables one to utilise the DiD approach whilst controlling for measured confounders. A similar approach could be taken in the analyses here; indeed, in the appendix we have generalised the results in the main paper to allow for covariate adjustment. A full study of the different estimators that may be available in this setting - such as likelihood-based, weighting-based and doubly robust methods - forms an interesting topic for additional work. 

\bibliographystyle{apalike}
\bibliography{bibfile.bib}

\appendix 

\section{Justification of approaches}

\subsection{Difference in differences}

In what follows, we will give results for a general data structure 
$\{Y_0,L_1,A,R_1,R_1Y_1,...,,L_T,R_T,R_TY_T\}$, where $\bar{L}_T=\{L_1,...,L_T\}$ is a time-varying variable that satisfies the conditional parallel trends assumption:
\begin{align}\label{cpt}
E(Y_T-Y_{t-1}|\bar{R}_{t}=1,\bar{L}_t,A=a)=E(Y_T-Y_{t-1}|R_{t}=0,\bar{R}_{t-1}=1,\bar{L}_t,A=a)
\end{align}
for all $t=1,...,T$ and $a=0,1$. Then we will show how assumption (\ref{cpt}) can be used in order to identify $E(Y_T^a)$. In addition to (\ref{cpt}), we will invoke the following assumptions that are standard in the causal inference/missing data literature:
\begin{itemize}
    \item (Consistency) If $A=a$, then $Y=Y^a$.
    \item (Positivity) $Pr(A=a|L_1=l_1)>0$ for all $a$ and $l_1$ where $f(l_1)\neq 0$. Also  if $Pr(R_t=1|\bar{R}_{t-1}=1,\bar{L}_t=\bar{l}_t,A=a)>0$ for all $a$ and $l_t$ where $f(\bar{R}_{t-1}=1,l_t,a)\neq 0$, at each $t=1,..,T$.
    \item (Conditional exchangeability) $A\indep Y^a|L_1=l_1 \forall a$.
\end{itemize}
These results allow for $A$ to be non-randomised, so long as the baseline covariates $L_1$ are sufficient to account for confounding.

Using a decomposition of the conditional mean $E(Y_T|\bar{R}_{T}=1,\bar{L}_T=\bar{l}_T,A=a)$ similar to that in 
\citet{tchetgen2017general}, we have
\begin{align*}
&E(Y_T|\bar{R}_{T}=1,\bar{L}_T=\bar{l}_T,A=a)\\
&=\{E(Y_T|\bar{R}_{T}=1,\bar{L}_T=\bar{l}_T,A=a)-E(Y_T|R_{T}=0,\bar{R}_{T-1}=1,\bar{L}_T=\bar{l}_T,A=a)\}\{1-\pi_{T}(\bar{l}_T)\}\\
&\quad +\delta_T(\bar{l}_T)-E\{\delta_T(\bar{L}_T)|\bar{R}_{T-1}=1,\bar{L}_{T-1}=\bar{l}_{T-1}\}\\
&\quad+E(Y_T|\bar{R}_{T-1}=1,\bar{L}_{T-1}=\bar{l}_{T-1},A=a)
\end{align*}
where we use the notation
\begin{align*}
\pi_t(\bar{L}_t)&=Pr(R_t=1|\bar{R}_{t-1}=1,\bar{L}_t,A=a)\\
\delta_t(\bar{l}_t)&=E(Y_T|\bar{R}_{t-1}=1,\bar{L}_t=l_t,A=a)-E(Y_T|\bar{R}_{t-1}=1,L_t=0,\bar{L}_{T-1}=\bar{l}_{T-1},A=a)
\end{align*}
Note that (\ref{cpt}) implies that 
\begin{align*}
&E(Y_T|\bar{R}_{T}=1,\bar{L}_T=\bar{l}_T,A=a)\\
&=\{E(Y_{T-1}|\bar{R}_{T}=1,\bar{L}_T=\bar{l}_T,A=a)-E(Y_{T-1}|R_{T}=0,\bar{R}_{T-1}=1,\bar{L}_T=\bar{l}_T,A=a)\}\{1-\pi_{T}(\bar{l}_T)\}\\
&\quad +\delta_T(\bar{l}_T)-E\{\delta_T(\bar{L}_T)|\bar{R}_{T-1}=1,\bar{L}_{T-1}=\bar{l}_{T-1}\}\\
&\quad+E(Y_T|\bar{R}_{T-1}=1,\bar{L}_{T-1}=\bar{l}_{T-1},A=a)
\end{align*}
and one can use a similar expansion for $E(Y_T|\bar{R}_{T-1}=1,\bar{L}_{T-1}=\bar{l}_{T-1},A=a)$. Working backwards in time down to $t=1$, we have
\begin{align*}
&E(Y_T|R_1=1,\bar{L}_1=\bar{l}_1,A=a)\\
&=\{E(Y_0|R_1=1,L_1=l_1,A=a)-E(Y_0|R_1=0,L_1=l_1,A=a)\}\{1-\pi_1(l_1)\}\\
&\quad+E(Y_T|L_1=l_1,A=a)
\end{align*}
Then adding all the preceding terms in the series together and re-arranging gives
\begin{align*}
&E(Y_T|L_1=l_1,A=a)\\
&=E(Y_T|\bar{R}_{T}=1,\bar{L}_T=\bar{l}_T,A=a)\\
&\quad -\sum^{T}_{t=1}\{E(Y_{t-1}|\bar{R}_t=1,\bar{L}_t=\bar{l}_t,A=a)-E(Y_{t-1}|R_t=0,\bar{R}_{t-1}=1,\bar{L}_t=\bar{l}_t,A=a)\}\{1-\pi_{t}(\bar{l}_t)\}\\
&\quad  -\sum^{T}_{t=2}[\delta_t(\bar{l}_t)-E\{\delta_t(\bar{L}_t)|\bar{R}_{t-1}=1,\bar{L}_{T-1}=\bar{l}_{T-1},A=a\}]
\end{align*}
where one can verify that the right hand side depends only on the observed data. Under the standard structural assumptions given above, $E(Y_T^a)$ can then be identified as 
\[E(Y_T^a)=E\{E(Y_T|L_1,A=a)\}\]
\citep{robins1986new}.

\subsection{Bespoke instrumental variables} 

Consider now a general data structure 
$\{Y_0,L_1,Z_1,A,R_1,R_1Y_1,...,,L_T,Z_T,R_T,R_TY_T\}$, where
$\bar{Z}_t$ satisfies the following restrictions:
\begin{itemize}
    \item (Stability of association) For all $t=0,...,T$ and $a=0,1$,
    \begin{align}
    &E(Y_T-Y_{t-1}|\bar{R}_{t-1}=1,Z_{t}=1,\bar{Z}_{t-1},\bar{L}_t,A=a)\nonumber\\
    &=E(Y_T-Y_{t-1}|\bar{R}_{t-1}=1,Z_{t}=0,\bar{Z}_{t-1},\bar{L}_t,A=a)\label{cer_l}
    \end{align}
    \item (Bias homogeneity) For all $t=0,...,T$ and $a=0,1$,
    \begin{align}\label{cnsm_l}
    &E(Y_T-Y_{t-1}|\bar{R}_{t}=1,Z_{t}=1,\bar{Z}_{t-1},\bar{L}_t,A=a)\nonumber\\&\quad-
    E(Y_T-Y_{t-1}|R_{t}=0,\bar{R}_{t-1}=1,Z_{t}=1,\bar{Z}_{t-1},\bar{L}_t,A=a)\nonumber\\
    &=E(Y_T-Y_{t-1}|\bar{R}_{t}=1,Z_{t}=0,\bar{Z}_{t-1},\bar{L}_t,A=a)\nonumber\\&\quad-
    E(Y_T-Y_{t-1}|R_{t}=0,\bar{R}_{t-1}=1,Z_{t}=0,\bar{Z}_{t-1},\bar{L}_t,A=a)
    \end{align}
    \item (Predictive of missingness) For all $t=0,...,T$ and $a=0,1$,
    \begin{align}\label{cpos_l}
        &Pr(R_t=1|\bar{R}_{t-1}=1,Z_{t}=1,\bar{Z}_{t-1},\bar{L}_t,A=a)-Pr(R_t=1|\bar{R}_{t-1}=1,Z_{t}=0,\bar{Z}_{t-1},\bar{L}_t,A=a)\nonumber\\&\neq 0 \quad \textrm{ almost surely.}
    \end{align}\end{itemize}

Then using a similar expansion as in the previous proof, 
\begin{align*}
&E(Y_T-Y_{T-1}|\bar{R}_{T}=1,\bar{Z}_T=\bar{z}_T,\bar{L}_T=\bar{l}_T,A=a)\\
&=\{E(Y_T-Y_{T-1}|\bar{R}_{T}=1,\bar{Z}_T=\bar{z}_T,\bar{L}_T=\bar{l}_T,A=a)\\&\quad -E(Y_T-Y_{T-1}|R_{T}=0,\bar{R}_{T-1}=1,\bar{Z}_T=\bar{z}_T,\bar{L}_T=\bar{l}_T,A=a)\}\{1-\pi_{T}(\bar{z}_T,\bar{l}_T)\}\\
&\quad +E(Y_T-Y_{T-1}|\bar{R}_{T-1}=1,\bar{Z}_T=\bar{z}_T,\bar{L}_T=\bar{l}_T,A=a)\\
&=\{E(Y_T-Y_{T-1}|\bar{R}_{T}=1,\bar{Z}_{T-1}=\bar{z}_{T-1},\bar{L}_T=\bar{l}_T,A=a)\\&\quad-E(Y_T-Y_{T-1}|R_{T}=0,\bar{R}_{T-1}=1,\bar{Z}_{T-1}=\bar{z}_{T-1},\bar{L}_T=\bar{l}_T,A=a)\}\{1-\pi_{T}(\bar{z}_T,\bar{l}_T)\}\\
&\quad +E(Y_T-Y_{T-1}|\bar{R}_{T-1}=1,\bar{Z}_{T-1}=\bar{z}_{T-1},\bar{L}_T=\bar{l}_T,A=a)
\end{align*}
by application of (\ref{cer_l})-(\ref{cpos_l}). As in \citet{tchetgen2017general}, this suffices to identify $E(Y_T-Y_{T-1}|\bar{R}_{T-1}=1,\bar{Z}_{T-1}=\bar{z}_{T-1},\bar{L}_T=\bar{l}_T,A=a)$, and therefore $E(Y_T|\bar{R}_{T-1}=1,\bar{Z}_{T-1}=\bar{z}_{T-1},\bar{L}_T=\bar{l}_T,A=a)$. This is because for fixed $\bar{L}_T$ and $\bar{Z}_{T-1}$,
$E(Y_T-Y_{T-1}|\bar{R}_{T}=1,\bar{Z}_T=\bar{z}_T,\bar{L}_T=\bar{l}_T,A=a)$ has two degrees of freedom, as does the right hand side of the second equality above. Note that $\pi_{T}(\bar{z}_T,\bar{l}_T)$ is considered as known here. Further,
\begin{align*}
&E(Y_T-Y_{T-1}|\bar{R}_{T-1}=1,\bar{Z}_{T-1}=\bar{z}_{T-1},\bar{L}_T=\bar{l}_T,A=a)\\
&=\delta^*_{T,T-1}(\bar{z}_{T-1},\bar{l}_T)-E\{\delta^*_{T,T-1}(\bar{z}_{T-1},\bar{L}_T)|\bar{R}_{T-1}=1,\bar{Z}_{T-1}=\bar{z}_{T-1},\bar{L}_{T-1}=\bar{l}_{T-1},A=a\}\\
&\quad+E(Y_T-Y_{T-1}|\bar{R}_{T-1}=1,\bar{Z}_{T-1}=\bar{z}_{T-1},\bar{L}_{T-1}=\bar{l}_{T-1},A=a)
\end{align*}
where
\begin{align*}
\delta^*_{T,t}(\bar{z}_{T-1},\bar{l}_t)=&E(Y_T-Y_{t}|\bar{R}_{t-1}=1,\bar{Z}_{T-1}=\bar{z}_{T-1},\bar{L}_t=l_t,A=a)\\&-E(Y_T-Y_{t}|\bar{R}_{t-1}=1,\bar{Z}_{T-1}=\bar{z}_{T-1},L_t=0,\bar{L}_{T-1}=\bar{l}_{T-1},A=a)
\end{align*}

One can then extend this reasoning to identify $E(Y_T|\bar{R}_{t}=1,\bar{Z}_{t}=\bar{z}_{t},\bar{L}_t=\bar{l}_t,A=a)$ for all $t=2,...,T$. At $t=1$, we have 
\begin{align*}
&E(Y_T-Y_0|R_1=1,Z_1=z_1,L_1=l_1,A=a)\\
&=\{E(Y_T-Y_0|R_1=1,L_1=l_1,A=a)-E(Y_T-Y_0|R_1=0,L_1=l_1,A=a)\}\\ &\quad\times \{1-\pi_1(z_1,l_1)\}\\
&\quad +E(Y_T-Y_0|L_1=l_1,A=a)
\end{align*}
and by combining all previous terms 
\begin{align*}
&E(Y_0|L_1=l_1,A=A)+\sum^{T}_{t=1}E(Y_{t}-Y_{t-1}|\bar{R}_t=1,\bar{Z}_t=\bar{z}_t,\bar{L}_t=\bar{l}_t,A=a)\\
&+\sum^{T}_{t=1}\delta^*_{T,t}(\bar{z}_{T-1},\bar{l}_t)-E\{\delta^*_{T,t}(\bar{z}_{T-1},\bar{L}_t)|\bar{R}_{T-1}=1,\bar{Z}_{T-1}=\bar{z}_{T-1},\bar{L}_{T-1}=\bar{l}_{T-1},A=a\}\\
&=E(Y_T|L_1=l_1,A=a)\\
&\quad+\sum^{T}_{t=1}\{E(Y_{t}-Y_{t-1}|\bar{R}_t=1,\bar{Z}_{t-1}=\bar{z}_{t-1},\bar{L}_{t}=\bar{l}_{t},A=a)\\&\quad-E(Y_{t}-Y_{t-1}|R_t=0,\bar{R}_{t-1}=1,\bar{Z}_{t-1}=\bar{z}_{t-1},\bar{L}_{t}=\bar{l}_{t},A=1)\}\pi_{t}(\bar{z}_t,\bar{l}_t)
\end{align*}
which one can leverage to obtain $E(Y_T|L_1=l_1,A=a)$. Under the structural assumptions in the previous subsection, this can be used in order to identify $E(Y^a_T)$.


\end{document}